\documentclass[11pt,draftcls,onecolumn]{article}
\usepackage{multicol,caption}
\usepackage{multirow}

\usepackage[a4paper, total={7in, 9in}]{geometry}
\usepackage{graphicx}
\usepackage{changepage}
\usepackage{enumitem}
\usepackage{listings}
\usepackage{amsmath}
\newenvironment{Figure}
{\par\medskip\noindent\minipage{\linewidth}}
{\endminipage\par\medskip}

\newcommand{\bc}{\begin{center}}
\newcommand{\ec}{\end{center}}
\newcommand{\tab}[1][1cm]{\hspace*{#1}}

\begin{document}
	\title{Informal Labour in India}
	\author{Vinay Reddy Venumuddala\\ PhD Student in Public Policy\\ Indian Institute of Management, Bangalore.}
	\date{}
	\maketitle

\begin{abstract}
		India like many other developing countries is characterized by huge proportion of informal labour in its total workforce. The percentage of Informal Workforce is close to 92\% of total as computed from NSSO 68th round on Employment and Unemployment, 2011-12\textsuperscript{\cite{nsso68}}. There are many traditional and geographical factors which might have been responsible for this staggering proportion of Informality in our country. \\
		\tab As a part of this study, we focus mainly on finding out how Informality varies with Region, Sector, Gender, Social Group, and Working Age Groups. Further we look at how Total Inequality\footnote{For the purpose of measuring inequality, we use Monthly Per-capita Consumption Expenditure} is contributed by Formal and Informal Labour, and how much do occupations/industries contribute to inequality within each of formal and informal labour groups separately. For the purposes of our study we use NSSO rounds 61 (2004-05) and 68 (2011-12) on employment and unemployment. The study intends to look at an overall picture of Informality, and based on the data highlight any inferences which are visible from the data.
\end{abstract} 

\section{Introduction}
	The term \textit{Informal Sector} in Indian context is more or less synonymous to \textit{Unorganized sector}, which signifies those enterprises that are outside purview of a majority of labour laws\textsuperscript{\cite{nceus_stats}}. Therefore workforce in these enterprises lack the protection of fair standards that are essential for enabling decent forms of work. These fair labour standards may include protection in terms of Social Security, Employment Security, Occupational Safety, freedom of collective bargaining, absence of any forms of discrimination at work and so on. In addition to workforce in Informal Sector, there are workforce in Formal Sector and also in households who often lack similar protection.	Since majority of workforce in the country are informal in nature, in a way it reflects the extent to which labour laws are ensuring protection to the total workforce. \\ \\
	\tab From a legal standpoint, an important notion that is widely used in countries around the world to assess the scope and extent of labour standards applicable between employee and employer is the \textit{Employment Relationship}. It is the legal interpretation attributed to Employment Relationship that guides the level of enforcement of labour standards. Following is the formal definition as described by an ILO report on Employment Relationship\textsuperscript{\cite{EmpRel}}.\\ 
	\tab \textit{``The employment relationship is a legal notion widely used in countries around the world to refer to the relationship between a person called an employee (frequently referred to as a worker) and an employer for whom the employee performs work under certain conditions in return for remuneration.\textsuperscript{\cite{EmpRel}}''} \\ \\
	\tab With globalization and growing use of Information Technology, the nature of working arrangements has become enormously diverse, and hence it poses a challenge in terms of the applicability of an Employment Relationship. More often, certain work arrangements like self-employed, triangular work relationships, and disguised employment restricts or completely nullifies the applicability of an employment relationship. To improve the extent of labour protection offered under law, it is therefore necessary to clarify the legal nature of employment relationship to cover most of the Occupations and Industries.\\ \\
	 \tab In order to approach the problem of Informality from a legal perspective, it is essential to look at overall nature to understand the complexity involved. With this as the motivation we try to look at the variation of informality across different categories (like Occupation, Industries, Gender, Soical Group, Sector, Region, etc..,) so that it can give a picture of the complexity we are dealing with. We also look at the contribution of formal and informal employment to Total Inequality and further within formal and informal groups, how different occupations or industries are contributing to these within group (formal/informal) inequalities.\\ \\
	\tab In Section-I we briefly give the definitional aspect of Informal labour, In Section-II we describe the classification within each of the categories in our study, In Sections-III and IV, we deal with variation of informality within categories, and the inequality analysis as described above. In Section-V we conclude by highlighting the limitations of this study and try to suggest possible areas of future work in this area. 

\section{Section-I: Definition of Informal Employment}
In its report on Definitional and Statistical Issues relating to Informal Economy\textsuperscript{\cite{nceus_stats}}, NCEUS has given clear and concise definitions for \textit{Informal Sector} and \textit{Informal Employment} in the context of India. In the following sub-section we give a brief overview of the definitional aspects of these terms. 
\subsection{Informal Sector and Informal Employment}
In order to arrive at the definition of Informal Sector, the commission has investigated three relevant major characteristics in line with international recommendations. 
\begin{enumerate}
	\item \textit{Enterprises owned by individuals or households that are not constituted as separate legal entities independent of their owners:} Committee has carefully investigated into data sources on unorganized enterprises (56th and 57th rounds of NSSO on unorganized manufacturing and service sector enterprises, 3rd Census of Small Scale Industries and 5th Economic Census on unorganized sector) and found out that more than 95\% of enterprises not covered under Organized Sector Surveys and Statistics (Like Annual Survey of Industries for organized manufacturing, National Accounts Statistics, etc..,) are proprietary and partnership in nature. 
	\item \textit{Employment Size of the enterprise has to be below a certain threshold to be determined according to national circumstances:}The committee in its report has studied various labour legislations offering protection in terms of social security, and income security to its workers, and has shown that the enterprises where number of workers are less than 10 are essentially not covered by any of the existing labour laws.
	\item \textit{Non-maintenance of complete accounts that would permit financial separation of production activities:} Committee identified that all the proprietary and partnership enterprises employing less than 10 total workers are not under any legal obligation to maintain separate accounts.
\end{enumerate}

Based on observing the characteristics of enterprises compatible with international recommendations, the committee has given the following definition of informal sector, \\

\textit{``The informal sector consists of all unincorporated private enterprises owned by individuals or households engaged in the sale and production of goods and services operated on a proprietary or partnership basis\footnote{An enterprise is classified as proprietary if an individual is its sole owner and as partnership if there are two or more owners on a partnership basis with or without formal registration. Partnerships also include SHGs and associations of individuals which are not recognized as separate legal entities. It excludes all corporate entities, cooperatives, trusts and other legal entities which are independent of their owners.} and with less than ten total workers''}\\

In addition to employees working in Informal Enterprises, there is informal employment also present in households and also in some formal sector enterprises. In order to cover these, the committee adopted the conceptual framework on Informal Employment, recommended in 17th ICLS, and has defined informal workers as follows, \\

\textit{``\textbf{Informal Employment:} Informal workers consist of those working in the informal sector or households, excluding regular workers with social security benefits provided by the employers and the workers in the formal sector without any employment and social security benefits provided by the employers''.} \\

In our analysis in subsequent sections we therefore categorize the workforce into formal and informal groups, using the above definition of Informal employment given by NCEUS\textsuperscript{\cite{nceus_stats}}.

\section{Section-II: Classification within categories}
In order to analyze the variation of Informality, we adopt the following major classifications so as to look at how different is the level of informality within each of these categories. We use Occupation, Industry, Sector (Rural/Urban), Region, Gender (Male/Female), Social Group (ST,SC,OBC and Others), and Working Age Group as our categories of interest. Following tables show the classification in some of the major categories we use during our data analysis in the further sections. \\
\begin{Figure}
	\captionsetup{font=scriptsize}
	\begin{center}
		\includegraphics[width=6.0in]{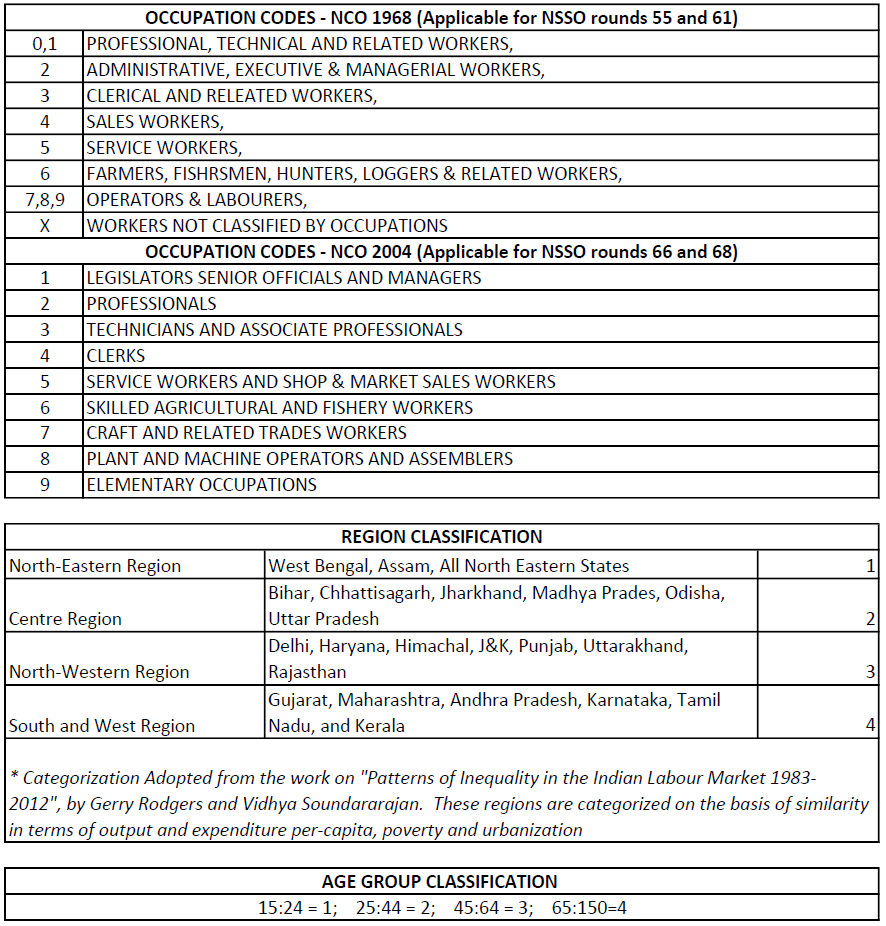}
		\captionof{figure}{Occupation Codes, Region Classification and Age-Group Classification} 
		\label{fig:PovOcc}
	\end{center}	
\end{Figure}

\begin{Figure}
	\captionsetup{font=scriptsize}
	\begin{center}
		\includegraphics[width=7.0in]{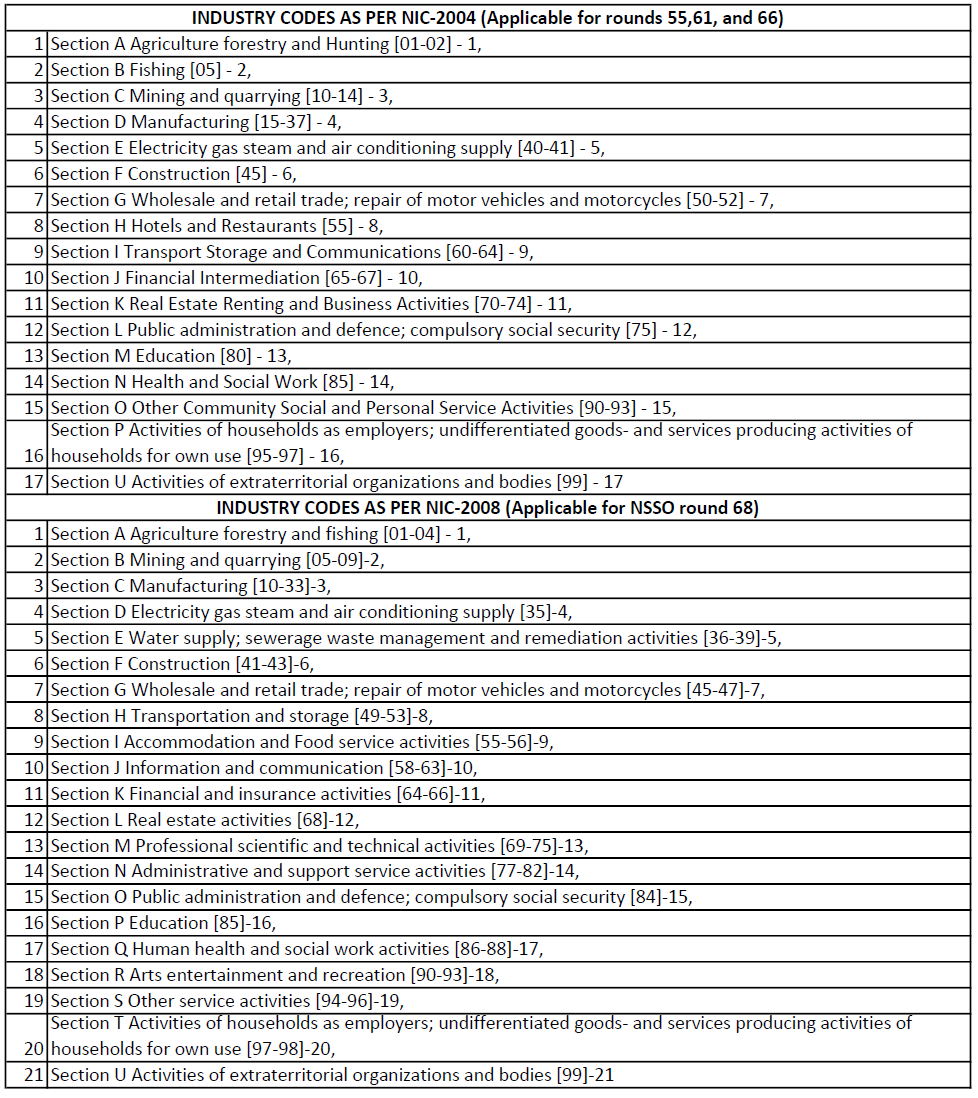}
		\captionof{figure}{Industry Codes (as per NIC-2004, and NIC-2008)} 
		\label{fig:IndCodes}
	\end{center}	
\end{Figure}

\section{Section-III: Analysis of Informality}

\subsection{Informality across Occupations and Industries}
In the following tables below, we give an insight into the percentage of Informal labour within and across each of the various industries and occupations in the country computed from NSSO 68th round data on Employment and Unemployment\textsuperscript{\cite{nsso68}}. 
\begin{Figure}
	\captionsetup{font=scriptsize}
	\begin{center}
		\includegraphics[width=7.0in]{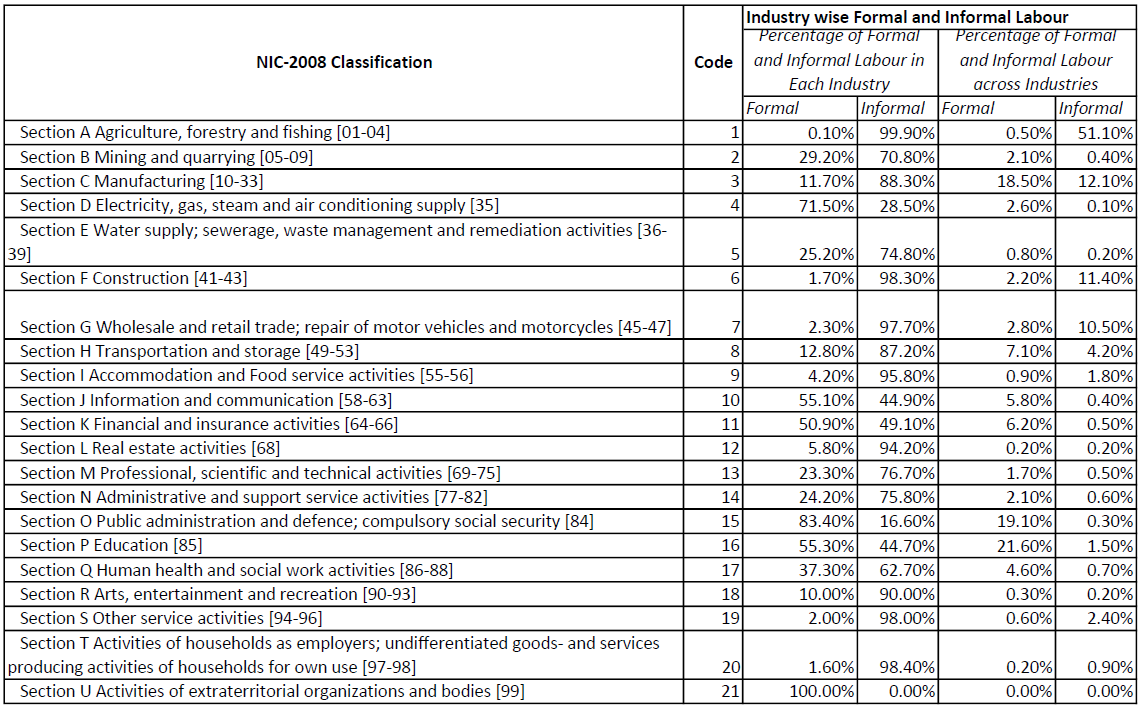}
		\captionof{figure}{Informality Pattern in Industries 2011-12} 
		\label{fig:Ind68}
	\end{center}	
\end{Figure}

\begin{Figure}
	\captionsetup{font=scriptsize}
	\begin{center}
		\includegraphics[width=6.0in]{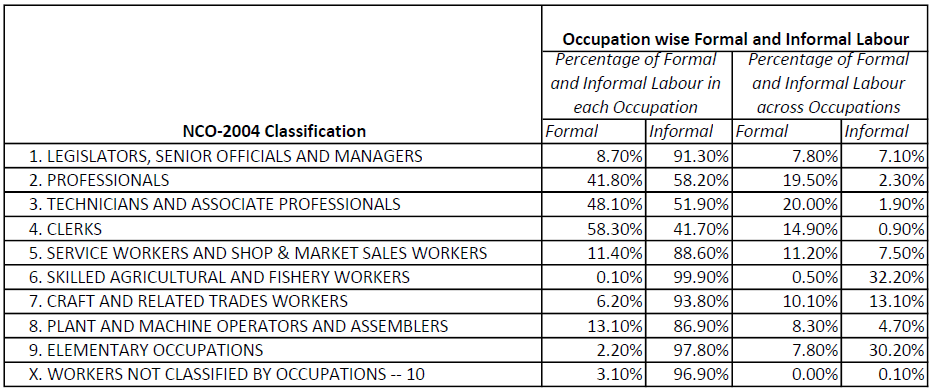}
		\captionof{figure}{Informality Pattern in Occupations 2011-12} 
		\label{fig:Occ68}
	\end{center}	
\end{Figure}

From the tables\footnote{First two columns in Figure\ref{fig:Ind68} shows the percentage of formal labour and informal labour within each industries, while the second set of two columns indicate the overall percentage of formal or informal labour across all the industries (Note that the total percentages in Column3 and in Column4 each add up to 100\%). Interpretation is similar for Figure \ref{fig:Occ68}} above we can see that Industry codes ..,9,6,19,20,7,12,1 contribute to higher proportion of informal labour within them, compared to 15,4,16,11,17,... Former include sectors like `Accommodation and Food Service Activities', `Construction', `Wholesale and retail trade and repair of motorcycles', `Real estate activities' and primarily `Agriculture, forestry and fishing'. While the latter include sectors like `Public	Administration, defense and social security', `Electricity, Gas, and Steam Supply', `Education', `Financial and Insurance activities', `Human health and social work', and so on. \\ 
\tab Similarly, occupational classes ..,7,9,1,6 have higher proportion of informal labour within them, than 4,3,2,... Former include classes like `Craft and related trades workers', `Skilled agriculture and fishery occupations', `Elementary occupations', and more importantly it also includes `Senior officials and managers'. Latter include classes like `Technicians', `Professionals', `Clerks', and so on. \\ \\
Above data clearly shows that the proportion of workforce beyond the coverage of labour laws are varying with occupation and industries. As we mentioned previously about the legal perspective of looking at fair labour standards, we can say that the employment relationship clarified by various labour laws in the country is somehow leaving out varying proportion of workers in different occupations and industries. The point we try to convey here is that, while determining the factors and indicators\footnote{Factors are the criteria used in determining the employment relationship, and indicators are used to identify whether or not relevant factors are present which determine the existence of an employment relationship.\textsuperscript{\cite{EmpRel}}} in describing employment relationship under labour legislations, it is essential to look at the differences in terms of work relationships that might arise in various industrial and occupational categories. Only when a comprehensive analysis of this sort is carried out, one can look into the applicability of fair labour standards across the country covering a majority of its workforce.

\subsection{Variation of Informality within categories}
In this sub-section we consider occupations as a way of looking at total informal workforce. And, we show how across occupations proportion of informality is varying with Working Age Groups, Regions, Sector, Gender, and Social Groups. We use NSSO 68th round on Employment and Unemployment\textsuperscript{\cite{nsso68}} to show the variation of Informality within categories.  
\begin{multicols}{2}
\begin{Figure}
	\captionsetup{font=scriptsize}
	\begin{center}
		\includegraphics[width=2.5in]{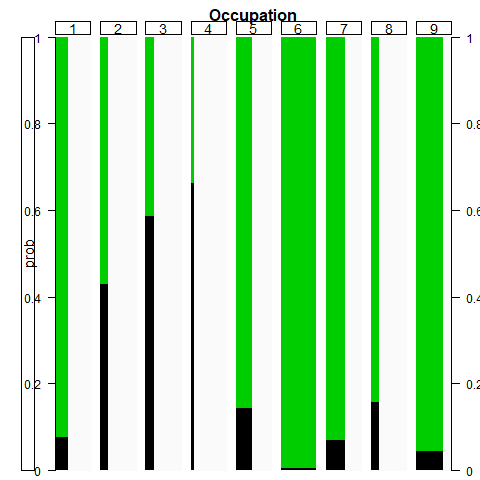}
		\captionof{figure}{Informality in various Occupation groups - 68th round} 
		\label{fig:OccOverall}
	\end{center}	
\end{Figure}

\begin{Figure}
	\captionsetup{font=scriptsize}
	\begin{center}
		\includegraphics[width=2.5in]{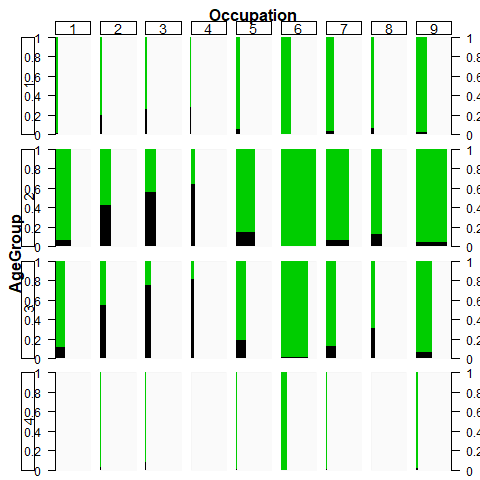}
		\captionof{figure}{Informality in various Occupation groups across age groups - 68th round} 
		\label{fig:OccAgeGroups}
	\end{center}	
\end{Figure}

\begin{Figure}
	\captionsetup{font=scriptsize}
	\begin{center}
		\includegraphics[width=3.0in]{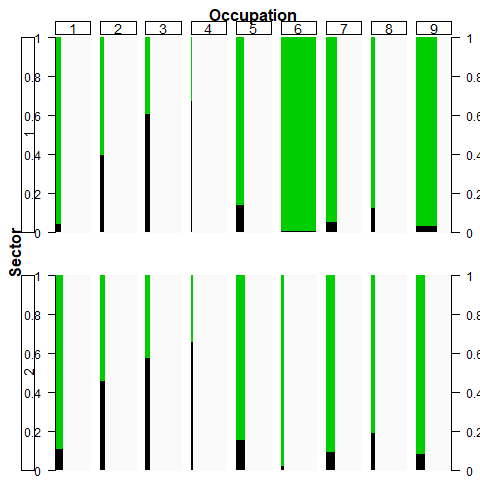}
		\captionof{figure}{Informality in various Occupation groups across Sector - 68th round (1- Rural, 2- Urban)} 
		\label{fig:OccSector}
	\end{center}	
\end{Figure}

\begin{Figure}
	\captionsetup{font=scriptsize}
	\begin{center}
		\includegraphics[width=3.0in]{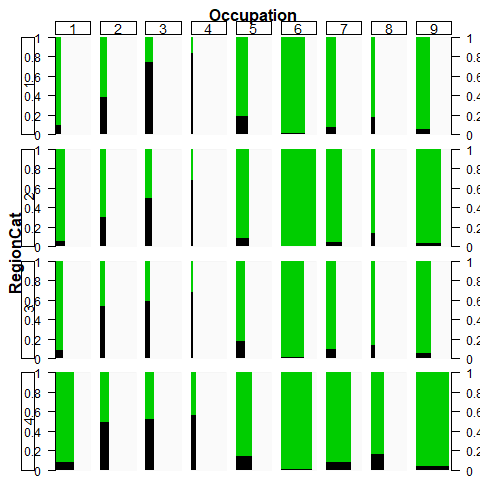}
		\captionof{figure}{Informality in various Occupation groups across Region Categories - 68th round} 
		\label{fig:OccRegionCat}
	\end{center}	
\end{Figure}

\begin{Figure}
	\captionsetup{font=scriptsize}
	\begin{center}
		\includegraphics[width=3.0in]{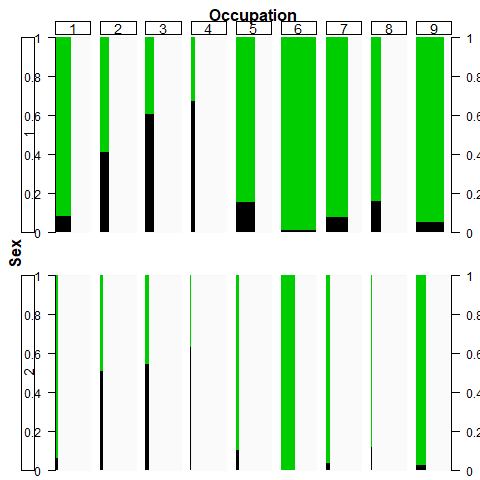}
		\captionof{figure}{Informality in various Occupation groups across Gender - 68th round (1- Male, 2- Female)} 
		\label{fig:OccGender}
	\end{center}	
\end{Figure}

\begin{Figure}
	\captionsetup{font=scriptsize}
	\begin{center}
		\includegraphics[width=3.0in]{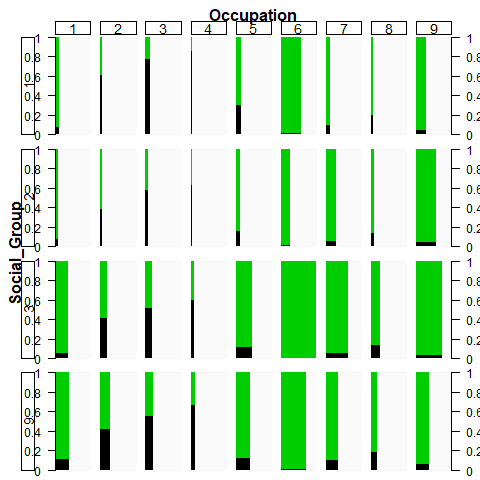}
		\captionof{figure}{Informality in various Occupation groups across Social Group (1-ST, 2-SC, 3- OBC, 9-Others) - 68th round} 
		\label{fig:OccSocialGroups}
	\end{center}	
\end{Figure}
\end{multicols}
In the above figures, proportion of informal labour is shown in green while proportion of formal labour is shown in black. And, thickness of each bar represents the number of workforce in respective occupations. Following are some inferences we can get from looking at the above plots.
\setlist{nolistsep}
\begin{itemize}[noitemsep]
	\item From Figure-\ref{fig:OccAgeGroups}, we can see that Working Age Groups 1 (15yrs-24yrs) and 4 (above 65yrs) account for higher proportion of informality in almost all the occupational classes.
	\item From Figure-\ref{fig:OccSector}, we can see that the proportion of informal labour working in Occupational codes 6 \textit{(Skilled Agricultural and fishery workers)}, and 9 \textit{Elementary Occupations} are differing significantly between Rural and Urban areas. This is obvious because of the predominance of agriculture workforce in villages. 
	\item Apart from number of labour working in different occupational classes, there is no clear visible variation (from just looking at the plots) in proportion of informality within categories of regions, Gender and Social Group. Deeper analysis is required in this context to identify if there are any trends in terms of variation of informality in certain occupational classes within each of these categories. 
\end{itemize}

\section{Section-IV: Inequality Decomposition on the basis of Informality}

\subsection{Inequality Decomposition - Theory}

We use decomposition by population subgroups as explained by Cowell et.al\textsubscript{\cite{cowell2011inequality}}. Following portion is a summary of the method taken from the paper \\ \\
Consider a Data Generating Process (DGP) represented in the following linear form:
\begin{align*}
	Y &= \beta_0+\sum_{k=1}^{K}\beta_k X_k + U \\
	Where,  \ \textbf{X}: &= (X_1,..,X_k) \ are \ explanatory \ random \ variables
\end{align*}
Now consider $X_1$ is a discrete random variable that take only the values ${X_{1,j}:j = 1,...,t_1}$. Data Generating Process (or population regression equation) for Sub-Group $j$ can be represented as:
\begin{align*}
Y_j &= \beta_{0,j}+\beta_{1,j} X_{1,j} + \sum_{k=2}^{K}\beta_{k,j} X_{k,j} + U_j \\
\end{align*}
Define $P_j = Pr(X_1 = X_{1,j})$, the proportion of population for which $X_1 = X_{1,j}$. Then within group inequality can be written as:
\begin{align*}
I_w(Y) = \sum_{j=1}^{t_1}W_j I(Y_j)
\end{align*}
where $t_1$ is the number of groups considered, $W_j$ is a weight that is a function of the $P_j$. The decomposition by population subgroups can be written as:
\begin{align*}
I(Y) = I_b(Y) + I_w(Y)
\end{align*}
where $I_b$ is the between-group inequality. In the case of Generalized Entropy (GE) Indices we have, for any $\alpha \in (-\infty,\infty)$,
\begin{align*}
W_j = P_j\left[ \frac{\mu(Y_j)}{\mu(Y)} \right]^{\alpha} = R_j^{\alpha} P_j^{1-\alpha},
\end{align*}
where $R_j := P_j \mu(Y_j)/\mu(Y)$ is the income share of group $j$, $\mu(Y_j)$ is mean income for subgroup $j$, $\mu(Y)$ is the mean income for the whole population. \\ 
\subsection{Empirical Application}
Our empirical application of this inequality decomposition proceeds as follows:
\begin{enumerate}
	\item We divide the individual data into subgroups by the variable Informality. Relating to the above theory, the random variable representing formal and informal groups can be given as\\ \\
	\textbf{Inequality Decomposition of Total Employed Workforce by Formal or Informal Groups:}
	\begin{align*}
	X_{1,j}:j=0,1 \;\ \ 0 = Formal, \ 1=Informal
	\end{align*}
	\begin{align*}
	Y_j = \beta_{0,j}+\beta_{1,j} X_{1,j} + \sum_{k=2}^{K}\beta_{k,j} X_{k,j} + U_j 
	\end{align*}
	We then compute the within group and between group inequality using $\alpha = 1.3$. Following are the equations of the total inequality in employed workforce, and decomposition.
	\begin{align*}
	I(Y) &= \frac{1}{\alpha^2 - \alpha}\left[\int \left[ \frac{Y}{\mu(Y)} \right]^\alpha dF(Y) - 1\right] \\
	I(Y) &= I_b(Y) + I_w(Y)\\
	I_w(Y) &= W^fI(Y^f) + W^iI(Y^i)\\
	I_b(Y) &= I(Y) - I_w(Y) \\
	W^f &= P^f\left[ \frac{\mu(Y^f)}{\mu(Y)} \right]^{\alpha} = (R^f)^{\alpha} (P^f)^{1-\alpha} \\
	W^i &= P^i\left[ \frac{\mu(Y^i)}{\mu(Y)} \right]^{\alpha} = (R^i)^{\alpha} (P^i)^{1-\alpha}
	\end{align*}
	Contributions to Within are given by
	\begin{align*}
	C_w(Formal) &= W^fI(Y^f) \\
	C_w(Informal) &= W^iI(Y^i) \\
	\end{align*}
	Contributions to Total is given by
	\begin{align*}
	C_t(Formal) &= 100\left[\frac{W^fI(Y^f)}{I(Y)} \right]  \\
	C_t(Informal) &= 100\left[\frac{W^iI(Y^i)}{I(Y)} \right]  \\
	\end{align*}
	\item In this step, we subset the individual data on the basis of informality. We have therefore two sub-sets, a. Individuals who are informally employed, and b. Individuals who are formally employed. Now, we separately compute the within group and between group inequality on each of these data subsets by decomposing on the basis of occupation codes. The random variable(s) and the sub-group DGP pertaining to these sub-sets can be given by \\
	
	\textbf{Inequality Decomposition of Formal Workforce by Occupational Codes:} 
	\begin{align*}
	X^f_{1,j} :j=1,2,..9; \ (Occupational \ Codes)
	\end{align*}
	\begin{align*}
	Y^f_j = \beta^f_{0,j}+\beta^f_{1,j} X^f_{1,j} + \sum_{k=2}^{K}\beta^f_{k,j} X^f_{k,j} + U^f_j 
	\end{align*}
		\tab Compute the within group and between group inequality using $\alpha = 1.3$. Following are the equations of the inequality in formal workforce, and further decomposition.
	\begin{align*}
	I(Y^f) &= I_b(Y^f) + I_w(Y^f)\\
	I_w(Y^f) &= \sum_{j=1}^{9}W^f_jI(Y^f_j) \\
	I_b(Y^f) &= I(Y^f) - I_w(Y^f)\\
	W^f_j &= P^f_j\left[ \frac{\mu(Y^f_j)}{\mu(Y^f)} \right]^{\alpha} = (R^f_j)^{\alpha} (P^f_j)^{1-\alpha} 
	\end{align*}
	Contributions to Within are given by
	\begin{align*}
	C_w(Each \ Occupation) &= W^f_jI(Y^f_j) \\
	\end{align*}
	Contributions to Total is given by
	\begin{align*}
	C_t(Each \ Occupation) &= 100\left[\frac{W^fW^f_jI(Y^f_j)}{I(Y)} \right]  \\
	\end{align*}

	\textbf{Inequality Decomposition of Informal Workforce by Occupational Codes:}
	\begin{align*}
	X^i_{1,j} :j=1,2,..9; \ (Occupational \ Codes) 
	\end{align*}
	\begin{align*}
	Y^i_j = \beta^i_{0,j}+\beta^i_{1,j} X^i_{1,j} + \sum_{k=2}^{K}\beta^i_{k,j} X^i_{k,j} + U^i_j 
	\end{align*}
		\tab Compute the within group and between group inequality using $\alpha = 1.3$. Following are the equations of the inequality in informal workforce, and further decomposition.
	\begin{align*}
	I(Y^i) &= I_b(Y^i) + I_w(Y^i)\\
	I_w(Y^i) &= \sum_{j=1}^{9}W_jI(Y^i_j) \\
	I_b(Y^i) &= I(Y^i) - I_w(Y^i) \\
	W^i_j &= P^i_j\left[ \frac{\mu(Y^i_j)}{\mu(Y^i)} \right]^{\alpha} = (R^i_j)^{\alpha} (P^i_j)^{1-\alpha}
	\end{align*}
	Contributions to Within are given by
	\begin{align*}
	C_w(Each \ Occupation) &= W^i_jI(Y^i_j) \\
	\end{align*}
	Contributions to Total is given by
	\begin{align*}
	C_t(Each \ Occupation) &= 100\left[\frac{W^iW^i_jI(Y^i_j)}{I(Y)} \right]  \\
	\end{align*}
\end{enumerate}

\begin{table}[ht!]
	\centering
	\begin{tabular}{llllllll}
		\hline
		& $ C_w $ & $ GEI $ & $ P $ & $ R $ & W/B & Index & $ C_t $(\%) \\ 
		\hline
		\multicolumn{7}{c}{\textbf{\textit{Decomposition by Formal/Informal Groups}}} \\
		\hline
		$ I(Y) $ & - & - & - & - & - & 0.282 & 100 \\ 
		\hspace{0.1cm}$ I_w(Y) $ & - & - & - & - & 0.241 & - & 85.29 \\ 
		\hspace{0.1cm}$ I_b(Y) $ & - & - & - & - & 0.041 & - & 14.71 \\ 
		\hspace{0.3cm}Formal & 0.056 & 0.275 ($ I(Y^f) $) & 0.08 & 0.165 & - & - & 20 \\ 
		\hspace{0.3cm}Informal & 0.184 & 0.227 ($ I(Y^i) $) & 0.92 & 0.835 & - & - & 65.29 \\ 
		\hline
		\multicolumn{7}{c}{\textbf{\textit{Decomposition of Formal Labour by Occupational Groups}}} \\
		\hline
		$ I(Y^f) $ & - & - & - & - & - & 0.275 & - \\ 
		\hspace{0.1cm}$ I_w(Y^f) $ & - & - & - & - & 0.229 & - & 16.65 \\ 
		\hspace{0.1cm}$ I_b(Y^f) $ & - & - & - & - & 0.046 & - & 3.35 \\ 
		\hspace{0.3cm}(F)Mgrs & 0.039 & 0.251 ($ I(Y^f_1) $) & 0.078 & 0.134 & - & - & 2.86 \\ 
		\hspace{0.3cm}(F)Professionals & 0.064 & 0.223 ($ I(Y^f_2) $) & 0.195 & 0.264 & - & - & 4.69 \\ 
		\hspace{0.3cm}(F)Technicians & 0.048 & 0.252 ($ I(Y^f_3) $) & 0.201 & 0.192 & - & - & 3.47 \\ 
		\hspace{0.3cm}(F)Clerks & 0.025 & 0.189 ($ I(Y^f_4) $) & 0.149 & 0.136 & - & - & 1.83 \\ 
		\hspace{0.3cm}(F)Service and Sales & 0.017 & 0.201 ($ I(Y^f_5) $) & 0.112 & 0.09 & - & - & 1.23 \\ 
		\hspace{0.3cm}(F)Agri and Allied & 0 & 0.106 ($ I(Y^f_6) $) & 0.005 & 0.003 & - & - & 0.02 \\ 
		\hspace{0.3cm}(F)Craft related & 0.009 & 0.14 ($ I(Y^f_7) $) & 0.101 & 0.073 & - & - & 0.68 \\ 
		\hspace{0.3cm}(F)Machine Op. & 0.011 & 0.198 ($ I(Y^f_8) $) & 0.083 & 0.062 & - & - & 0.81 \\ 
		\hspace{0.3cm}(F)Elem. Occ & 0.014 & 0.361 ($ I(Y^f_9) $) & 0.078 & 0.047 & - & - & 1.05 \\ 
		\hline
		\multicolumn{7}{c}{\textbf{\textit{Decomposition of Informal Labour by Occupational Groups}}} \\
		\hline
		$ I(Y^i) $ & - & - & - & - & - & 0.223 & - \\ 
		\hspace{0.1cm}$ I_w(Y^i) $ & - & - & - & - & 0.194 & - & 55.68 \\ 
		\hspace{0.1cm}$ I_b(Y^i) $ & - & - & - & - & 0.033 & - & 9.61 \\ 
		\hspace{0.3cm}(I)Mgrs & 0.037 & 0.283 ($ I(Y^i_1) $) & 0.071 & 0.113 & - & - & 10.59 \\ 
		\hspace{0.3cm}(I)Professionals & 0.018 & 0.35 ($ I(Y^i_2) $) & 0.024 & 0.043 & - & - & 5.23 \\ 
		\hspace{0.3cm}(I)Technicians & 0.009 & 0.247 ($ I(Y^i_3) $) & 0.019 & 0.03 & - & - & 2.45 \\ 
		\hspace{0.3cm}(I)Clerks & 0.003 & 0.178 ($ I(Y^i_4) $) & 0.009 & 0.015 & - & - & 0.87 \\ 
		\hspace{0.3cm}(I)Service and Sales & 0.018 & 0.191 ($ I(Y^i_5) $) & 0.075 & 0.089 & - & - & 5.13 \\ 
		\hspace{0.3cm}(I)Agri and Allied & 0.05 & 0.177 ($ I(Y^i_6) $) & 0.322 & 0.29 & - & - & 14.28 \\ 
		\hspace{0.3cm}(I)Craft related & 0.02 & 0.16 ($ I(Y^i_7) $) & 0.13 & 0.127 & - & - & 5.77 \\ 
		\hspace{0.3cm}(I)Machine Op. & 0.008 & 0.14 ($ I(Y^i_8) $) & 0.047 & 0.056 & - & - & 2.35 \\ 
		\hspace{0.3cm}(I)Elem. Occ & 0.031 & 0.142 ($ I(Y^i_9) $) & 0.302 & 0.237 & - & - & 9 \\ 
		\hline
	\end{tabular}
	\caption{\label{Decomp} Inequality Decomposition (Computed from NSSO Round 68)}
\end{table}

\section{Section-V: Discussion}
\tab Informal Labour is a major policy issue, which particularly the developing countries are facing. To tackle the problem of informality, it is crucial to look at it holistically, before formulating broad policy solutions. Above study is just a small step towards understanding the problem of Informality in India based on recent NSSO rounds on Employment and Unemployment. It is limited by the extent to which it highlights the nature of informality in the country, and we see that there is a lot that needs to be researched in this area. Further work can be done to look within each of the industry and occupational classes and see how categories like Gender, Social Group, Sector, Working Age Groups etc.., are interacting with them to impact the extent of informality. It is essential to understand the problem of informality from the view-point of all the relevant stakeholders involved, so that formulating policies can be streamlined properly. \\
\tab Further there is a need to look from a legal perspective into each of the industrial and occupational classes and find if the scope of employment relationship can be improved in our labour laws. We conclude by saying that the above analysis is only one of the many perspectives of looking at how a country can realize fair labour standards for its workforce. Even after expanding the scope and extent of employment relationship, many forms of work arrangements where the workforce is Self-Employed or Casual Labour are still going to be outside this purview. Therefore it is important to carry out further research into how a country can accommodate labour of this nature within the definition of its employment relationship, or may be find out how it can legally enable fair labour standards to these workers outside the traditional framework of an employment relationship.

\section{Appendix}

\bibliographystyle{Generic}
\bibliography{reflibrary}

\begin{thebibliography}{1}
\expandafter\ifx\csname url\endcsname\relax
  \def\url#1{\texttt{#1}}\fi
\expandafter\ifx\csname urlprefix\endcsname\relax\def\urlprefix{URL }\fi
\expandafter\ifx\csname href\endcsname\relax
  \def\href#1#2{#2} \def\path#1{#1}\fi

\bibitem{nsso68}
NSSO, India - employment and unemployment, july 2011- june 2012, nss 68th
  round.

\bibitem{nceus_stats}
NCEUS, Report on definitional and statistical issues relating to informal
  economy (2008) Chapter 2: Concepts and Definitions.

\bibitem{EmpRel}
ILO, The employment relationship, report v1, international labour organization,
  geneva, ILO.

\bibitem{cowell2011inequality}
F.~A. Cowell, C.~V. Fiorio, Inequality decompositions—a reconciliation,
  Journal of Economic Inequality 9~(4) (2011) 509--528.

\end{thebibliography}

\end{document}